\documentclass[prd,aps,twocolumn]{revtex4}
\usepackage{graphicx}
\usepackage{epstopdf}
\usepackage{amssymb}
\usepackage{color}
\usepackage{natbib}
\usepackage{xspace}

\def\xmm{XMM-Newton\xspace}

\begin{document}
\title{Constraints on 3.55~keV line emission from stacked observations of dwarf spheroidal galaxies}
\author{D.~Malyshev}
\author{A.~Neronov}
\author{D.~Eckert}
\affiliation{Department of Astronomy, University of Geneva, ch. d'Ecogia 16, CH-1290 Versoix, Switzerland}
\begin{abstract}
Several recent works have reported the detection of an unidentified X-ray line at 3.55~keV, which could possibly be attributed to the decay of dark matter (DM) particles in the halos of galaxy clusters and in the M31 galaxy. We analyze all publicly-available \xmm data of dwarf spheroidal galaxies to test the possible DM origin of the line. Dwarf spheroidal galaxies have high mass-to-light ratios and their interstellar medium is not a source of diffuse X-ray emission, thus they are expected to provide the cleanest DM decay line signal. Our analysis shows no evidence for the presence of the line in the stacked spectra of the dwarf galaxies. It excludes the sterile neutrino DM decay origin of the 3.5 keV line reported by Bulbul et al. (2014) at the level of {\bf $4.1\sigma$} under standard assumptions about the Galactic DM column density in the direction of selected dwarf galaxies and at the level of {\bf $3.2\sigma$} assuming minimal Galactic DM column density.  Our analysis is still consistent with the estimate of  sterile neutrino DM parameters by Boyarsky et al. (2014), because of its larger uncertainty. However, the central value of their estimate of the mixing angle is inconsistent with our dwarf spheroidals data at $3.4\sigma\ (2.5\sigma)$ level assuming the mean (minimal) Galactic DM column density. As a by-product of our analysis, we provide updated upper limits to the mixing angle of sterile neutrino DM in the mass range between 2 and 20 keV.
\end{abstract}
\maketitle
\section{Introduction}
In a recent work,~\citet{bulbul:14,boyarsky:14} reported the detection of an unidentified 3.55~keV line in the analysis of stacked data on galaxy clusters observed by \xmm, in the  observations of individual galaxy clusters (Perseus) and  in the Andromeda galaxy. In the absence of obvious options for this line to be of instrumental/astrophysical origin (see, however, \citet{Jeltema:14}), the authors invoked the possibility for the line to be produced in the decay of Dark Matter (DM) particles populating the halos of the considered structures. Assuming a sterile-neutrino nature of the decaying DM (see e.g. \citet{Dodelson:93,Asaka:05a,Lattanzi:07} and references therein), \citet{bulbul:14} obtained a value of $\sin^2(2\theta)\sim6.8\times 10^{-11}$ for the the mixing angle of the sterile neutrino and a particle mass $m_{DM}$ of 7.1~keV. These parameters are consistent with the previously-derived upper bounds on $\sin^2(2\theta)$ from observations of the extragalactic diffuse X-ray background~\cite{Boyarsky:05,Abazajian:06b};
galaxy clusters~\cite{Boyarsky:06b,Riemer-Sorensen:06b,Boyarsky:06e}; the Milky Way, Andromeda (M31) and Triangulum (M33) galaxies~\cite{Abazajian:06b,Watson:06,Boyarsky:06b,Boyarsky:06c,Boyarsky:06d,Boyarsky:07b,Yuksel:07} and individual dwarf spheroidal (dSph) satellites of the Milky Way
\cite{Boyarsky:06c,Boyarsky:06f,Boyarsky:10b,Riemer-Sorensen:09,Loewenstein:09,Loewenstein:12,Kusenko:12}.

The signal from decaying DM is expected to be strongest from the most nearby source, our own Milky Way galaxy. However, the signal is distributed over the entire sky, so that it is not straightforward to look for this signal using narrow-field X-ray telescopes. Besides, the radial density profile of DM in the Milky Way is somewhat uncertain, especially in its central part, where a significant contribution to the overall matter content of the Galaxy comes from the baryons. Recently \citet{riemer:14} analyzed Chandra data on the Galactic Center (GC) region, finding no clear evidence for the 3.55~keV line (see however \citet{Boyarsky:14b}). The non-detection of the signal from the GC was found to be consistent with the existence of the DM sterile neutrino with parameters suggested by \citet{bulbul:14} for the most conservative assumptions on the DM density profile in the innermost part of the Galaxy.  

The signal from the Milky Way halo is superimposed on approximately equally strong signal from the DM halos of nearby dSph galaxies in the direction of these sources. Combining the two signals and additionally stacking the signal from all the observed dSph systems, an improved sensitivity to the DM decay line can be obtained compared to the previously-reported constraints from the Milky Way or individual dSph galaxies. 

In the following, we perform a stacked analysis of dSph galaxies using \xmm\ data. In spite of a shorter overall exposure compared to the stacked galaxy cluster dataset analyzed by \citet{bulbul:14}, the signal from the dSph is cleaner than in galaxy clusters, because the interstellar medium of dSph galaxies is not a source of thermal X-ray emission, contrary to the intracluster medium of the galaxy clusters.  The work presented here is organized as follows. In the next section, we discuss the expected flux level from the dSph galaxies (Sect.~\ref{sec:dsphs_signal}) and the contribution of the Milky Way (MW) halo to the observed signal in Sect.\ref{sec:mw_contribution}. The details of the \xmm observations and of the analysis procedure are given in Sect.~\ref{sec:observations}. Finally, the results are summarized and discussed in Sect.~\ref{sec:results_discussion}.

\section{Expected signal from DM decay}
\label{sec:expected_signal}
%%%%%%%%%%%%%%%%%%%%%%%%%%%%%%%%
\subsection{dSph galaxies}
\label{sec:dsphs_signal}
%%%%%%%%%%%%%%%%%%%%%%%%%%%%%%%%

The decay of DM particles of mass $m_{DM}$ results in a line at the energy $\epsilon=m_{DM}/2$ and flux
\begin{equation}
F=\frac{\Gamma M_{DM,FoV}}{4\pi d^2 m_{DM}}
\label{eq:flux}
\end{equation}
where $M_{DM,FoV}$ is the total DM mass within the field of view (FoV) of the telescope. In the case of sterile neutrinos, the radiative decay width $\Gamma$ can be written as \cite{Pal:82}
\begin{eqnarray}
\Gamma&=&\frac{9\alpha G_F^2}{256 \cdot 4\pi^4}\sin^2 2\theta_{DM} m_{DM}^5\nonumber\\
&\simeq& 1.7\times 10^{-28}\left[\frac{\theta_{DM}^2}{1.75\times 10^{-11}}\right]\left[\frac{m_{DM}}{7.1\mbox{ keV}}\right]^5\mbox{ s}^{-1}
\end{eqnarray}
where we have normalized the mixing angle $\theta_{DM}$ and $m_{DM}$ to the values reported in~\citet{bulbul:14}.
Substituting this expression into Eq.~\ref{eq:flux} one finds the flux from dwarf spheroidal in the field of view to be 
\begin{eqnarray}
\label{eq:flux_num}
F_{dSph}\simeq 2.4\times 10^{-7}\left[\frac{\theta_{DM}^2}{1.75\times 10^{-11}}\right]\left[\frac{m_{DM}}{7.1\mbox{ keV}}\right]^4\nonumber\\\left[\frac{d}{100\mbox{ kpc}}\right]^{-2}
\left[\frac{M_{DM,FoV}}{10^{7}M_\odot}\right]\frac{\mbox{ph}}{\mbox{cm}^2\mbox{s}}
\end{eqnarray}

Note, that the flux~(\ref{eq:flux_num}) does not depend on the exact DM distribution, but only on the total mass inside the telescope's FoV. The latter can be measured with greater accuracy than the measurement of a profile. The uncertainty in the expected DM decay flux~(\ref{eq:flux_num}) can be directly propagated from the uncertainties in $M_{DM}$ measurements.

In our work we consider the sample of dwarf spheroidal galaxies presented by~\citet{wolf:10}, where the authors obtained accurate estimates of the DM mass inside the half-light radius, i.e. the radius within which half of the total light is observed. The choice of this integration radius is motivated by the fact that this radius minimizes the uncertainty in the enclosed mass. In Table~\ref{tab:dsphs_params} we summarize the available data on dSphs for which archival \xmm data are available. These data (from~\citet{wolf:10}) quote the mass estimates $M_{1/2}$ within the half-light radius sphere, together with the uncertainties and the values of the half-light radius both in physical ($r_{1/2}$)  and angular ($\theta_{1/2}$) size. For comparison we also show the masses within the columns out to $r_{1/2}$ from the recent work \citet{geringer:14}, where available.

For a number of dSphs (e.g. Ursa Minor), the half-light radius exceeds the size of the \xmm FoV, $\theta_{FoV}\simeq 15'$. This implies that only a fraction of the DM decay signal is visible in a single \xmm pointing.
We take this into account for the estimation of the DM flux from these particular galaxies. Namely, we assume that the mass within the FoV scales as $M_{DM,FoV}\sim \theta_{FoV}$ for the sources with $\theta_{1/2}>\theta_{FoV}$, taking into account the fact that the velocity dispersion profiles of dSphs are flat, see e.g.~\citet{walker:07,walker:13} and references therein. The centres of the field-of-view for most observations are slightly displaced from the dSphs centre positions, so that the maximal angular distance from the centre is typically smaller than the full FoV size. For sources spanning the full FoV we restrict the analysis to the region $\theta_{max}=12'$ around the source centre.

%%%%%%%%%%%%%%%%%%%%%%%%%%%%%%%%
\begin{table}
\begin{tabular}{l|llll|ll}
\hline
Name & $d$ & $M_{1/2}$ & $r_{1/2}$ & $\theta_{1/2}$&$M_{col,1/2}$ \\
     & kpc & $10^7M_\odot$ & kpc & arcmin   & $10^7 M_\bigodot$       \\
\hline
Carina &$105\pm 2$&   $0.95^{+0.095}_{-0.09}$&   0.334 & 10.9 &$1.9_{-0.3}^{+0.4}$ \\
\hline
Draco & $76\pm 5$&     $2.11^{+0.31}_{-0.31}$& 0.291 & 13.2  &$3.9_{-0.5}^{+0.6}$\\
\hline
Fornax &$147\pm 3$&   $7.39^{+0.41}_{-0.36}$&  0.944 & 22.1&$6.7_{-0.3}^{+0.2}$ \\
\hline
Leo I &$254\pm 18$& $2.21^{+0.24}_{-0.24}$&   0.388 & 5.3 &$5.5_{-0.6}^{+0.8}$ \\
\hline
Ursa Minor & $77\pm   4$&   $5.56^{+0.79}_{-0.72}$&   0.588 & 26.3 &$3.3_{-0.5}^{+0.5}$\\
\hline
Ursa Major II &$32\pm  4$&       $0.79^{+0.56}_{-0.31}$&  0.184 & 19.8&$0.8_{-0.3}^{+0.6}$\\
\hline
Willman I& $38\pm 7$&        $0.04^{+0.02}_{-0.02}$&  0.033 & 3.0 & --  \\
\hline
NGC 185 &$616\pm 26$ &   $29.3^{+10.2}_{-7.7}$&  0.355 & 2.0 & --  \\
\hline
\end{tabular}

\caption{{Parameters of dwarf spheroidal galaxies considered in the analysis.  The distances to the dwarves($d$, kpc), half-light radii ($r_{1/2}$,kpc and $\theta_{1/2}$, arcmin) and the masses within half-light radii sphere ( $M_{1/2}, 10^{7}M_\bigodot$ ) are taken from \citet{wolf:10}. The column masses within half-light radii are from \citet{geringer:14}, see text for the details.}}
\label{tab:dsphs_params}
\end{table}
%%%%%%%%%%%%%%%%%%%%%%%%%%%%%%%%

%%%%%%%%%%%%%%%%%%%%%%%%%%%%%%%%%%%%%%%%%%%%%
\subsection{Milky Way}
\label{sec:mw_contribution}
%%%%%%%%%%%%%%%%%%%%%%%%%%%%%%%%%%%%%%%%%%%%%

The DM decay flux from the Milky Way halo is typically comparable to the flux from isolated distant sources, like dSph galaxies or galaxy clusters \cite{boyarsky:06}. The flux from DM decay in the Milky Way within the telescope field-of-view $\Omega_{FoV}$,
\begin{equation}
F=\frac{\Gamma\Omega_{FoV}{\cal S}}{4\pi m_{DM}}
\end{equation}
is determined by the column density of the DM
\begin{equation}
{\cal S}=\int\limits_{0}^{\infty}\rho_{DM}\left(\sqrt{r^2_{\bigodot}-2zr_\bigodot\cos\phi+z^2}\right)dz
\label{eq:S_MW}
\end{equation}
where $r_\bigodot=8.5$~kpc is the distance from the Sun to the centre of our Galaxy and the angle $\phi$ relates to the galactic coordinates $(l,b)$ as $\cos\phi=\cos b \cos l$.

For a source of angular size $\theta$, the contribution of the MW to the DM decay flux is 
\begin{eqnarray}
F_{MW}=1.1\times 10^{-6}\left[\frac{\theta}{\theta_{FoV}}\right]^2\left[\frac{S}{10^{22}\mbox{GeV/cm}^2}\right]\nonumber\\
\left[\frac{\theta_{DM}^2}{1.75\times 10^{-11}}\right]\left[\frac{m_{DM}}{7.1\mbox{ keV}}\right]^4\mbox{ph/cm}^2\mbox{s}
\label{eq:flux_MW}
\end{eqnarray}

To estimate the column density $S$ of Galactic DM in different directions, we adopt the models of the DM halo of the Milky Way discussed by~\citet{klypin:02}. \citet{klypin:02} have adopted the Navarro-Frenk-White (NFW, \citet{nfw}) profile
\begin{equation}
\rho_{NFW}(r)=\frac{\rho_s r_s^{3}}{r(r+r_s)^2}
\label{eq:nfw}
\end{equation}
in which the characteristic density $\rho_s$ and radius $r_s$ are free parameters estimated from the data. 

The uncertainty in the column density as well as in the radial DM density profile arises from the difficulty of disentangling contributions from the visible and DM components to the Galaxy rotation curve. In what follows we consider the ``favoured NFW'' model of \citet{klypin:02} for the estimate of the \emph{mean} Milky Way DM column density in the direction of individual dSphs (with $\rho_s=4.9\times 10^6$~M$_\bigodot$kpc$^{-3}$, $r_s=21.5$~kpc). To estimate how the uncertainty propagates to the limits on 
the mixing angle of sterile neutrino DM, we also consider the column densities of Galactic DM deduced from the  ``maximal disk model'' ($\rho_s=0.6\times 10^6$~M$_\bigodot$kpc$^{-3}$, $r_s=46$~kpc) of \citet{klypin:02}. We refer to this estimate as the \emph{minimal} Galactic DM contribution to the signal.  For each position in the sky the \emph{minimal} DM column density is  2-3~times lower than the estimated \emph{mean} column density.

The expected Galactic contribution to the DM decay flux from individual dSphs is given in Table \ref{tab:predicted_flux}. The total expected DM decay flux from  the direction of each dSph galaxy is the sum of the fluxes given in Eqs.  (\ref{eq:flux_num}) and (\ref{eq:flux_MW}). The expected fluxes $F_{dSph}$ are given in two column format for the mass estimations by \citet{wolf:10} and \citet{geringer:14} correspondingly.

%%%%%%%%%%%%%%%%%%%%%%%%%%%%%%%%%%%%%%%
\begin{table}
\begin{tabular}{|c|c|c|c|c|c|}
\hline
Object & Exposure,  & \multicolumn{2}{|c|}{F$_{dSph}$,} & F$_{MW, mean}$ & F$_{MW, min}$ \\
       & \tiny {$10^7$ cm$^2$s }  & \multicolumn{2}{|c|}{\tiny{$10^{-7}$cts/cm$^2$/s}}& \tiny{$10^{-7}$cts/cm$^2$/s} & \tiny{$10^{-7}$cts/cm$^2$/s} \\
\hline
Carina & 1.41 & $2.1^{+0.21}_{-0.20} $ & $ 4.1^{+0.9}_{-0.7} $ & 5.9 & 2.7   \\
Draco  & 6.23 & $8.0^{+1.2}_{-1.2}  $ & $ 14.7^{+2.3}_{-1.9} $& 8.1 & 3.6 \\
Fornax & 5.35 & $4.5^{+0.25}_{-0.22} $ & $ 4.0^{+0.12}_{-0.20}$ & 6.9 &  3.2 \\
Leo    & 5.03 & $0.8^{+0.09}_{-0.09} $ & $ 2.0^{+0.3}_{-0.2}  $ & 1.2 & 0.56 \\
NGC185 & 12.75& $1.9^{+0.66}_{-0.50} $  &  --& 0.4 & 0.15 \\
UMa II & 0.87 & $11.2^{+7.95}_{-4.4} $ & $ 11.4^{+8.5}_{-4.3} $ &  5.6 & 2.7 \\
UMi    & 0.86 & $10.3^{+1.45}_{-1.33} $ & $ 6.1^{+0.9}_{-0.9}$ & 7.0 &  3.2 \\
Willman& 8.36 & $0.7^{+0.35}_{-0.35}$  &  --  & 0.4 & 0.2 \\
\hline
TOTAL  & 40.86 &$3.17^{+0.71}_{-0.58} $ & $ 4.26^{+0.91}_{-0.7}$ & 3.0 & 1.4\\
\hline
\end{tabular}

\caption{Expected fluxes from DM decay line with parameters corresponding to \citet{bulbul:14}. For each dwarf spheroidal galaxy the expected flux is split into two components: the signal from the dwarf itself ($F_{dsph}$, see Eq.~\ref{eq:flux_num}) and from the Milky Way ($F_{MW}$, see Eq.~\ref{eq:flux_MW}). For the MW contribution flux estimates are given for both mean and minimal DM profiles.The two columns for F$_{dSph}$ correspond to the flux values from $M_{1/2}$ mass estimations by \citet{wolf:10} and \citet{geringer:14} correspondingly.}
\label{tab:predicted_flux}
\end{table}
%%%%%%%%%%%%%%%%%%%%%%%%%%%%%%%%%%%%%%%

%%%%%%%%%%%%%%%%%%%%%%%%%%%%%%%%%%%%%%%
\section{Observations and data analysis}
\label{sec:observations}
%%%%%%%%%%%%%%%%%%%%%%%%%%%%%%%%%%%%%%%

The details of the \xmm observations of the dSph galaxies selected for this analysis are summarized in Table~\ref{tab:observations}. We collected all publicly-available observations of eight dSphs with exposures exceeding 10~ks. The total exposure time of the observations is $\sim 0.6$~Msec. In most cases, the observations were already used for the search of the DM decay signal and upper bounds on the sterile-neutrino DM mixing angle were already derived on a source-by-source basis~\citep{Loewenstein:09,Riemer-Sorensen:09,Loewenstein:12}. The goal of our re-analysis of these data is to stack the signal from all the dSphs to increase our sensitivity to the DM decay line.

We processed the raw data with the ESAS (v.0.9.28 as part of XMM SAS v.13.5) reduction scheme \citep{Snowden:08}\footnote{See e.g. http://heasarc.gsfc.nasa.gov/docs/xmm/esas /esasimage/esasimage\_thread.html} using the calibration files from May 2014. Within this scheme, we produced cleaned event files by removing the periods of soft proton flares. After the flare removal, we performed a self-consistency check on the level of contribution of residual soft protons to the background flux, following \citet{leccardi:08}. This contribution is found to be less than $\sim~20\%$ in all observations. In any case, the spectrum of the soft protons is featureless, so our line search is unaffected by soft-proton contamination. 

We masked all the detected point sources in the FoV using the ESAS task \texttt{cheese}, such that only the signal from the extended DM halo of the dSphs is taken into account. We then extracted spectra and images of the extended emission using the ESAS tasks \texttt{mos-spectra} and \texttt{pn-spectra}. These tools use a collection of closed-filter data to estimate the local non X-ray background (NXB) by scaling the normalization of the closed-filter spectra to match the count rates measured in the unexposed corners of each of the three EPIC detectors. 

%%%%%%%%%%%%%%%%%%%%%%%%%%%%%%%%%%%%%%%
\begin{table}[t!]
\begin{tabular}{|c|c|c|c|}
\hline
Obs Id & Name & Duration, & Clean exposure, \\
       &      & ksec      & ksec \\
\hline
0200500201 &  Carina & 41.9 & 19.2+16.7+8.4 \\
0603190101 &  Draco & 19.0  & 17.5+17.9+14.3 \\
0603190201 &  Draco & 19.9 & 18.5+18.2+14.7 \\
0603190301 &  Draco & 17.7 & 12.2+12.6+6.3 \\
0603190501 &  Draco & 19.9 & 18.6+18.5+14.9 \\
0302500101 &  Fornax & 103.9 & 65.1+65.9+53.0 \\
0555870201 &  Leo & 94.0 & 75.4+77.1+0 \\
0652210101 &  NGC 185 & 123.5 & 91.4+96.2+66.7 \\
0650180201 &  UMa II & 34.3 & 11.7+12.5+7.4 \\
0301690401 &  UMi & 11.8 & 10.8+10.9+7.9 \\
0652810101 &  Willman & 29.3 & 15.0+19.0+9.5 \\
0652810301 &  Willman & 36.0 &  21.9+23.2+15.5 \\
0652810401 &  Willman & 36.2 & 27.5+28.5+16.2 \\
\hline
TOTAL      &          & 602.3 & 404.8+417.2+232.8\\
\hline
\end{tabular}

\caption{\xmm observations of dwarf spheroidals considered in this analysis. The total exposure time is given as the sum of effective exposures for the MOS1, MOS2 and pn cameras individually. The total clean exposure is $\sim 0.6$~Msec}
\label{tab:observations}
\end{table}
%%%%%%%%%%%%%%%%%%%%%%%%%%%%%%%%%%%%%%%

We stacked the spectra of all the dSph observations using the {\it addspec} routine, to obtain mean MOS1+MOS2 and pn spectra. We then fitted the resulting spectra in the 0.7-10~keV energy band with the sum of models representing the astrophysical and NXB contributions. We ignored the energy interval 1.2-1.8~keV from the fit, as this range is affected by the presence of strong and time-variable Si K$\alpha$ and Al K$\alpha$ fluorescence lines.  

Instead of subtracting the NXB spectra directly from the data, we modelled the NXB spectrum using a phenomenological model including all known fluorescence lines (see Appendix B of \citet{leccardi:08}) and added this model as an additive component to the fit. This method has the advantage of retaining the original statistics of the spectrum and allowing for possible variations of the NXB level, e.g. caused by soft protons. During the fitting procedure, we fixed the spectral shape of the particle-induced continuum to the values obtained from the closed-filter data, since this component is known to be stable with time \citep{leccardi:08}. On the other hand, we leave the normalizations of the instrumental lines free while fitting. In the case of the pn camera we also need to model the Ca line at $\sim 4.6$~keV. We note that no instrumental line is observed between 3 and 4 keV, thus this analysis is suitable for the detection of an additional emission line in this energy range. For the details of the analysis procedure, we refer the reader to \citet{Eckert:14}.

\begin{figure*}[t!]
\resizebox{0.9\hsize}{!}{\includegraphics[angle=270]{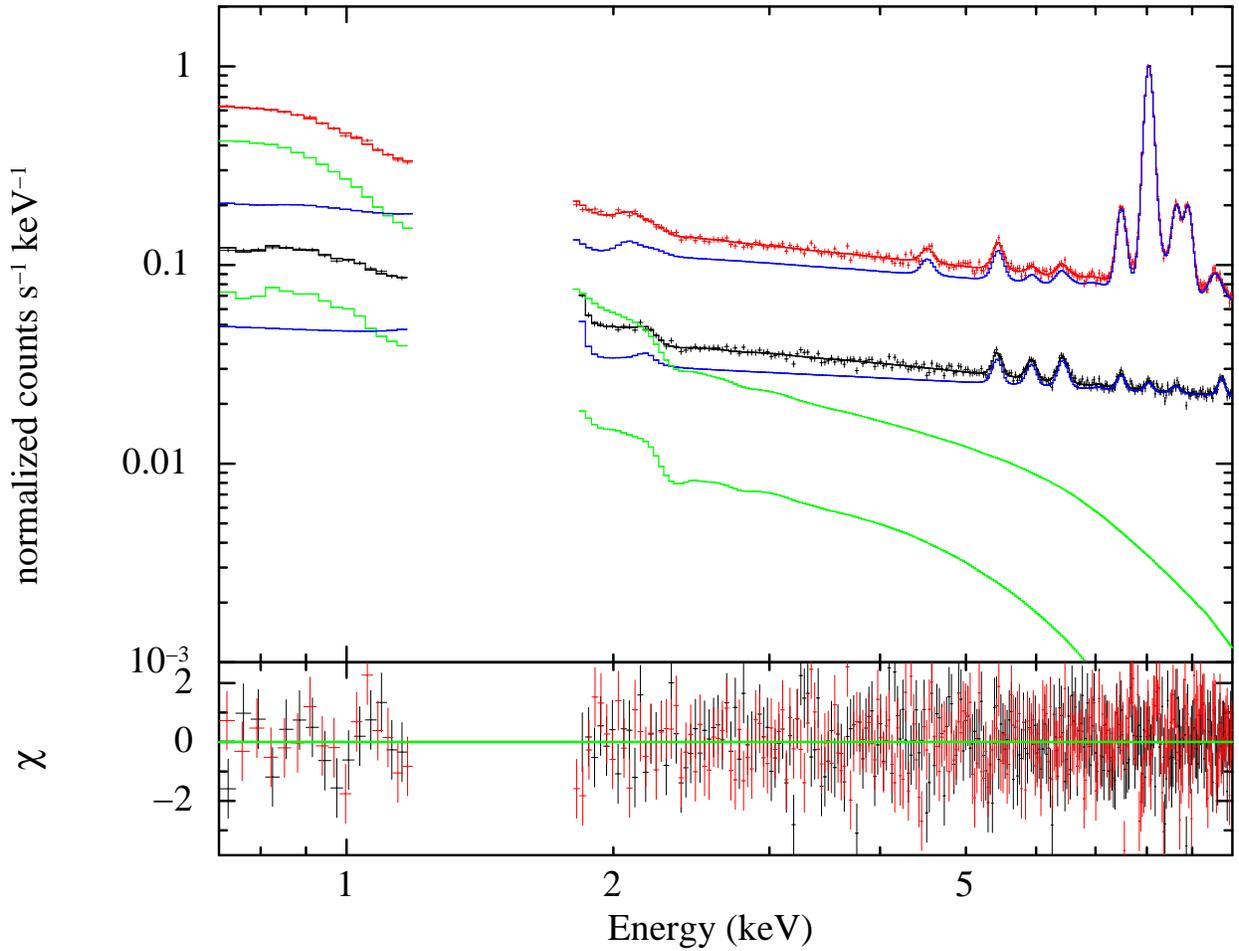}}
\vspace{0.6cm}
\caption{MOS1+2 (black) and pn (red) stacked spectra of dSph galaxies. The top panel shows the data and the best-fit model, broken into the astrophysical contribution (green) and the NXB (blue), while the bottom panel shows the residuals from the model.}
\label{fig:spectra}
\end{figure*}

\begin{figure}[t!]
\includegraphics[width=0.5\textwidth]{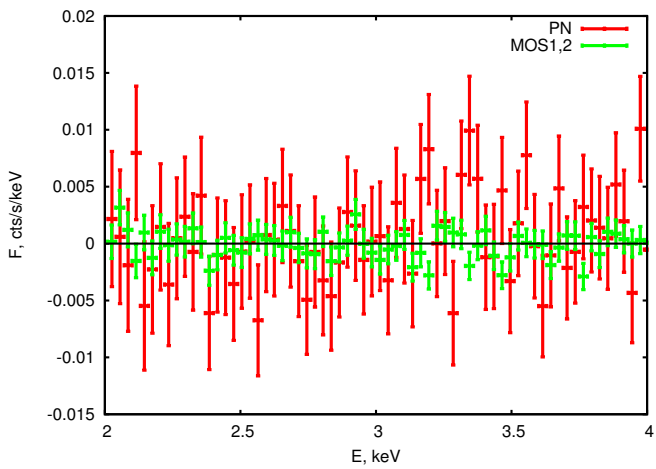}
\caption{Residual flux from the combined MOS~1,2  (green) and pn (red) cameras in the 2-4~keV energy band including the expected 3.55~keV line.}
\label{fig:residuals}
\end{figure}

The astrophysical background and foreground contribution is modelled as the sum of a power law representing the cosmic X-ray background (CXB) and two {\it apec} models for the Galactic emission. The combined fit is reasonably good, with $\chi^2/DOF=595.5/576$. In Fig. \ref{fig:spectra} we show the stacked spectra and best-fit model for pn (red) and MOS (black). The residuals from the fit are displayed in the bottom panel of the figures. The main parameters of the fit are given in Table \ref{tab:fit_params}. The fit requires the presence of a rather hot component at the temperature of $\sim0.9$ keV, which is significantly higher than the typical temperature \citep{Mccammon:02}; however the temperature of the Galactic halo is known to vary significantly from one direction to another, and thus such a result is not unusual \citep{Masui:09}. In any case, we note that the foreground emission from the MW halo is very soft compared to the energy range of interest for this study (see Fig. \ref{fig:spectra}), such that the exact temperature of the foreground component has little influence on our analysis. Moreover, the spectral slope of the CXB, which is the dominant sky component beyond $\sim1$ keV, is in excellent agreement with the canonical value of 1.4 \citep[e.g.][]{DeLuca:04}. The normalization of the CXB in each individual observation was also found to agree with the measurement of \citet{DeLuca:04}. A closer look at the residuals in the range of interest for this analysis, i.e. the range between 2 and 4 keV, is shown in Fig. \ref{fig:residuals}. 

%%%%%%%%%%%%%%%%%%%%%%%%%%%%%%%%%%%%%%%
\begin{table}[t!]
\begin{tabular}{|c|c|c|c|}
\hline
Model name & Parameter & Value & Error \\
\hline
  apec1  &     kT,keV  &    0.91    &  0.05   \\
%  apec1  &     Abundanc    &        1.00000   &   frozen \\
%  apec1   &    Redshift      &      0.0     &     frozen \\
  apec1  &     norm        &        3.5$\cdot 10^{-5}$  &  0.5$\cdot 10^{-5}$   \\
\hline
  apec2   &   kT, keV    &  0.35   &  0.06  \\
%  apec2   &    Abundanc      &      1.00000   &   frozen \\
%  apec2   &    Redshift     &       0.0       &   frozen \\
  apec2   &    norm        &        2.9$\cdot 10^{-5}$  &  0.7$\cdot 10^{-5}$  \\
\hline
  powerlaw &  PhoIndex     &       1.34      & 0.05  \\
  powerlaw  & norm         &       1.2$\cdot 10^{-4}$  &  6.0$\cdot 10^{-6}$ \\
\hline
\end{tabular}

\caption{Parameters of the fit of the stacked spectra of dSphs with the reference background model.}
\label{tab:fit_params}
\end{table}
%%%%%%%%%%%%%%%%%%%%%%%%%%%%%%%%%%%%%%%

%%%%%%%%%%%%%%%%%%%%%%%%%%%%%%%%%%%%%%%
%%%%%%%%%%%%%%%%%%%%%%%%%%%%%%%%%%%%%%%

%%%%%%%%%%%%%%%%%%%%%%%%%%%%%%%%%%%%%%%
%%%%%%%%%%%%%%%%%%%%%%%%%%%%%%%%%%%%%%%

\begin{figure}
\resizebox{0.9\hsize}{!}{\includegraphics[angle=270]{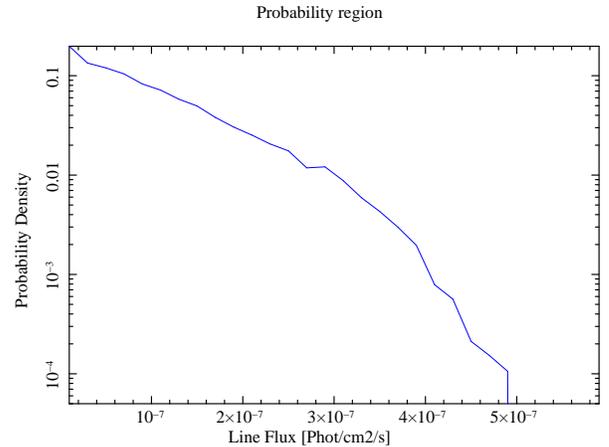}}
\vspace{0.2cm}
\caption{Posterior probability distribution for the 3.55 keV line flux in the stacked dSph dataset, obtained through 50,000 MCMC sampling of the likelihood function.}
\label{fig:mcmc}
\end{figure}

To search for the DM decay line, we added to the model a narrow gaussian line at a fixed energy of $3.55$~keV. The addition of such a line does not provide a significant improvement to the fit. To compute upper limits to the line flux marginalizing over all uncertainties, we sampled the likelihood using a Markov chain Monte Carlo (MCMC) method as implemented in XSPEC v12.8. After an initial burning phase of 5,000 steps, we performed 50,000 MCMC steps and drew the posterior distributions from the resulting chain. The output distribution for the line flux is shown in Fig. \ref{fig:mcmc}. From this distribution we obtained upper limits to the line flux of $2.97\times10^{-7}$ phot cm$^{-2}$ s$^{-1}$ (90\% confidence level) and $4.74\times10^{-7}$ phot cm$^{-2}$ s$^{-1}$ (3$\sigma$). In addition, we also performed a search for the DM decay line in the entire energy range 2-10~keV, which did not give any positive results. From this analysis we derived an energy-dependent upper bound on the line flux.

The contribution of the individual dSph galaxies to the \xmm\ signal could be calculated using the information on the instrument effective area $A_{i}$ (found using the {\it plot efficiency} command in XSPEC) and the clean exposure $t_i$ in each observation. If the expected flux of the DM decay line in the $i$th observation is $F_i$, the expected number of DM decay photons is $N_{i}=F_{i} A_i t_i$. The mean DM decay flux in the entire stacked dataset is then 
\begin{equation}
\left<F\right>=\frac{\sum\limits_i F_{i} A_i t_i} {\sum\limits_i A_i t_i },
\label{eq:Fmean}
\end{equation}
with the sum being performed over all dSphs observations. 

Table \ref{tab:predicted_flux} gives the information on $A_i$ and $t_i$ in each observation, together with the estimate of the expected DM decay line flux calculated assuming the sterile neutrino DM parameters suggested by the observations of \citet{bulbul:14}.

%%%%%%%%%%%%%%%%%%%%%%%%%%%%%%%%%%%%%%%
\section{Results and discussion}
\label{sec:results_discussion}
%%%%%%%%%%%%%%%%%%%%%%%%%%%%%%%%%%%%%%%

Our analysis of the stacked sample of dSph galaxies provides a moderate improvement of constraints on the parameters for sterile neutrino DM, compared to the previously-derived bounds based on the previous observations of the diffuse X-ray background, of individual dSph galaxies, of the Milky Way and of galaxy clusters. 

As it follows from Table~\ref{tab:predicted_flux}, the flux from DM decay expected to be seen in the combination of observations of all dwarves is $F_{mean}\sim 6.2\pm 0.7\times 10^{-7}$~cts/s/cm$^2$ for mean dark matter column density in the MW and $F_{min}\sim 4.6\pm 0.6\times 10^{-7}$~cts/s/cm$^2$ for minimal Galactic DM column density for DM mass estimates from \citet{wolf:10}. In the case of \citet{geringer:14} mass estimates the reference values are $F_{mean}\sim 7.3\pm 0.8\times 10^{-7}$~cts/s/cm$^2$ and $F_{min}\sim 5.7\pm 0.8\times 10^{-7}$~cts/s/cm$^2$. Such a flux was expected to be detected at  $3.5\sigma (4.1\sigma)$  or $2.8\sigma (3.2\sigma)$ levels for the mean / minimal Galactic DM column density models in the stacked dSph dataset for \citet{wolf:10} (\citet{geringer:14}) mass estimates. The non-detection of the DM line in the analyzed data sets is, therefore, inconsistent with the assumption that the unidentified line at 3.55~keV is produced by decaying sterile-neutrino DM with parameters suggested by the analysis of \citet{bulbul:14}. {

This is illustrated by Fig. \ref{fig:exclusion_plot}, where we plot the $2\sigma$ upper limits on the mixing angle of the DM sterile neutrino as a function of the DM particle mass, assuming quoted by \citet{geringer:14} dSphs masses. The value of $\sin^2(2\theta)$ derived by \citet{bulbul:14} is above the $2\sigma$ upper bound for both the \emph{mean} and \emph{minimal} Galactic DM column density models. 

%%%%%%%%%%%%%%%%%%%%%%%%%%%%%%%%%%%%%
\begin{figure}[t!]
\includegraphics[width=\linewidth]{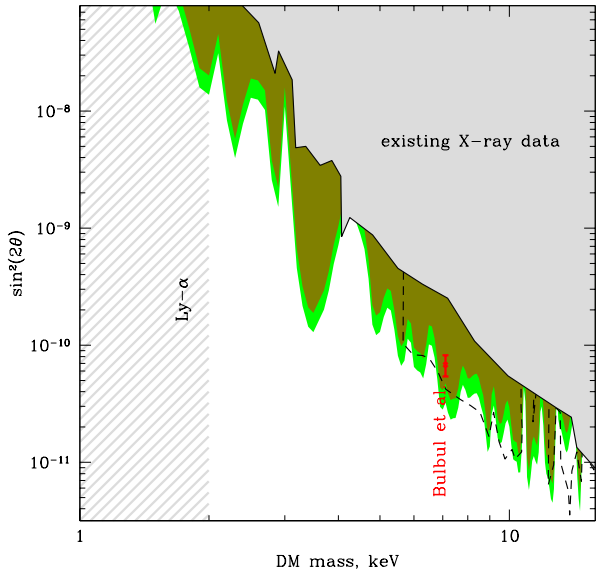}
\caption{Exclusion plot on sterile neutrino mass -- mixing angle plane. All parameter values above the curves are excluded. The solid dark green and light green lines show the $2\sigma$ constraints for minimal and mean dark matter column density in the Milky Way, respectively. The red point with error bars indicates the parameter values reported by \citet{bulbul:14}. The dashed line indicates the M31 constraints from \citet{horiuchi:14}.}
\label{fig:exclusion_plot}
\end{figure}
%%%%%%%%%%%%%%%%%%%%%%%%%%%%%%%%%%%%%

The calculation of the DM decay line flux from the direction of galaxy clusters by \citet{bulbul:14} does not include the flux from the foreground DM halo of the Milky Way. This is justified if the foreground flux is subtracted in the analysis of the spectra of the galaxy clusters. This is not the case in the analysis of \citet{bulbul:14}, who modelled the cluster spectra together with the instrumental and sky background / foreground in a way similar to the approach adopter here. In this case, the flux from DM decay in the MW halo should, in principle, be included in the calculation of the line flux \cite{boyarsky:06}. This should result in a somewhat lower value of the sterile neutrino mixing angle $\sin^2(2\theta)$ from the observed line flux. This effect might potentially relax the inconsistency of the \citet{bulbul:14} result with the dSph data reported here. 

The estimates of  $\sin^2(2\theta)$ derived by \citet{boyarsky:14} from the analysis of M31 and of the Perseus galaxy cluster are much more uncertain than those derived from the stacked galaxy cluster sample, because of the much larger uncertainty in the DM column density in the two particular individual sources. Taking into account a roughly order-of-magnitude uncertainty in the estimate of  $\sin^2(2\theta)$  derived from the analysis of \citet{boyarsky:14}, one could see that our constraints on the DM parameters derived from the dSph data are still consistent with the results of \citet{boyarsky:14}.

However, the central value of the estimate of $\sin^2(2\theta)\sim 5\cdot 10^{-11}$ from \citet{boyarsky:14} is inconsistent with the limits from dSphs at the level of $3.4\sigma (2.5\sigma)$ assuming the mean (minimal) Galactic DM column density and taking the column (rather than within-a-ball) mass within half-light radius estimates of \citet{geringer:14}.

Our analysis is only marginally ruling out the possibility of the DM decay origin of the unidentified line at 3.55~keV. 
An increase of the sensitivity by a factor of $\sim 2$ is necessary to firmly rule out the DM decay line hypothesis for the line origin. This is possible already with \xmm (rather than with the next-generation telescopes like ASTRO-H), via a moderate increase of exposure towards selected dSph galaxies (e.g. Ursa Minor, Ursa Major II), which are characterized by strong DM decay line flux, but are currently not dominating the stacked dSph signal because of the relatively short exposures. Deeper \xmm observations of these dSphs galaxies would thus be sufficient to test conclusively the DM origin of the 3.55 keV line.
\newline
\textit{Acknowledgements.} The authors would like to thank Alex Geringer-Sameth for providing the raw data from the recent analysis \citet{geringer:14}.

%%%%%%%%%%%%%%%%%%%%%%%%%%%%%%%%

% Bibliography and bibfile
\def\aj{AJ}%
          % Astronomical Journal
\def\actaa{Acta Astron.}%
          % Acta Astronomica
\def\araa{ARA\&A}%
          % Annual Review of Astron and Astrophys
\def\apj{ApJ}%
          % Astrophysical Journal
\def\apjl{ApJ}%
          % Astrophysical Journal, Letters
\def\apjs{ApJS}%
          % Astrophysical Journal, Supplement
\def\ao{Appl.~Opt.}%
          % Applied Optics
\def\apss{Ap\&SS}%
          % Astrophysics and Space Science
\def\aap{A\&A}%
          % Astronomy and Astrophysics
\def\aapr{A\&A~Rev.}%
          % Astronomy and Astrophysics Reviews
\def\aaps{A\&AS}%
          % Astronomy and Astrophysics, Supplement
\def\azh{AZh}%
          % Astronomicheskii Zhurnal
\def\baas{BAAS}%
          % Bulletin of the AAS
\def\bac{Bull. astr. Inst. Czechosl.}%
          % Bulletin of the Astronomical Institutes of Czechoslovakia
\def\caa{Chinese Astron. Astrophys.}%
          % Chinese Astronomy and Astrophysics
\def\cjaa{Chinese J. Astron. Astrophys.}%
          % Chinese Journal of Astronomy and Astrophysics
\def\icarus{Icarus}%
          % Icarus
\def\jcap{J. Cosmology Astropart. Phys.}%
          % Journal of Cosmology and Astroparticle Physics
\def\jrasc{JRASC}%
          % Journal of the RAS of Canada
\def\mnras{MNRAS}%
          % Monthly Notices of the RAS
\def\memras{MmRAS}%
          % Memoirs of the RAS
\def\na{New A}%
          % New Astronomy
\def\nar{New A Rev.}%
          % New Astronomy Review
\def\pasa{PASA}%
          % Publications of the Astron. Soc. of Australia
\def\pra{Phys.~Rev.~A}%
          % Physical Review A: General Physics
\def\prb{Phys.~Rev.~B}%
          % Physical Review B: Solid State
\def\prc{Phys.~Rev.~C}%
          % Physical Review C
\def\prd{Phys.~Rev.~D}%
          % Physical Review D
\def\pre{Phys.~Rev.~E}%
          % Physical Review E
\def\prl{Phys.~Rev.~Lett.}%
          % Physical Review Letters
\def\pasp{PASP}%
          % Publications of the ASP
\def\pasj{PASJ}%
          % Publications of the ASJ
\def\qjras{QJRAS}%
          % Quarterly Journal of the RAS
\def\rmxaa{Rev. Mexicana Astron. Astrofis.}%
          % Revista Mexicana de Astronomia y Astrofisica
\def\skytel{S\&T}%
          % Sky and Telescope
\def\solphys{Sol.~Phys.}%
          % Solar Physics
\def\sovast{Soviet~Ast.}%
          % Soviet Astronomy
\def\ssr{Space~Sci.~Rev.}%
          % Space Science Reviews
\def\zap{ZAp}%
          % Zeitschrift fuer Astrophysik
\def\nat{Nature}%
          % Nature
\def\iaucirc{IAU~Circ.}%
          % IAU Cirulars
\def\aplett{Astrophys.~Lett.}%
          % Astrophysics Letters
\def\apspr{Astrophys.~Space~Phys.~Res.}%
          % Astrophysics Space Physics Research
\def\bain{Bull.~Astron.~Inst.~Netherlands}%
          % Bulletin Astronomical Institute of the Netherlands
\def\fcp{Fund.~Cosmic~Phys.}%
          % Fundamental Cosmic Physics
\def\gca{Geochim.~Cosmochim.~Acta}%
          % Geochimica Cosmochimica Acta
\def\grl{Geophys.~Res.~Lett.}%
          % Geophysics Research Letters
\def\jcp{J.~Chem.~Phys.}%
          % Journal of Chemical Physics
\def\jgr{J.~Geophys.~Res.}%
          % Journal of Geophysics Research
\def\jqsrt{J.~Quant.~Spec.~Radiat.~Transf.}%
          % Journal of Quantitiative Spectroscopy and Radiative Trasfer
\def\memsai{Mem.~Soc.~Astron.~Italiana}%
          % Mem. Societa Astronomica Italiana
\def\nphysa{Nucl.~Phys.~A}%
          % Nuclear Physics A
\def\physrep{Phys.~Rep.}%
          % Physics Reports
\def\physscr{Phys.~Scr}%
          % Physica Scripta
\def\planss{Planet.~Space~Sci.}%
          % Planetary Space Science
\def\procspie{Proc.~SPIE}%
          % Proceedings of the SPIE
\let\astap=\aap
\let\apjlett=\apjl
\let\apjsupp=\apjs
\let\applopt=\ao
%\bibliography{all_lit}

\end{document}